\documentclass[twocolumn,showpacs,preprintnumbers,amsmath,amssymb]{revtex4}

\usepackage{graphicx}
\usepackage{dcolumn}
\usepackage{bm}

\begin{document}

\title{Comment on ``Scientific collaboration networks. II. Shortest paths, weighted networks, and centrality"}
\author{Tao Zhou$^{1}$}
\email{zhutou@ustc.edu}
\author{Jian-Guo Liu$^{2}$}
\author{Bing-Hong Wang$^{1}$}
\affiliation{%
$^{1}$Department of Modern Physics and Nonlinear Science Center,
University of Science and Technology of China, Hefei, 230026, PR
China\\
$^{2}$Institute of System Engineering, Dalian University of
Technology, Dalian, 116023, PR China
}%

\date{\today}

\begin{abstract}
In this comment, we investigate a common used algorithm proposed
by Newman [M. E. J. Newman, Phys. Rev. E {\bf 64}, 016132(2001)]
to calculate the betweenness centrality for all vertices. The
inaccurateness of Newman's algorithm is pointed out and a
corrected algorithm, also with O($MN$) time complexity, is given.
In addition, the comparison of calculating results for these two
algorithm aiming the protein interaction network of Yeast is
shown.
\end{abstract}


\maketitle

Betweenness centrality, also called load or betweenness for
simplicity, is a quite useful measure in the network analysis.
This conception is firstly proposed by
Anthonisse\cite{Anthonisse1971} and Freeman\cite{Freeman1977} and
introduced to physics community by Newman\cite{Newman2001}. The
betweenness of a node $v$ is defined as
\begin{equation}
B(v):=\sum_{s\neq t,s\neq v}\frac{\sigma_{st}(v)}{\sigma_{st}},
\end{equation}
where $\sigma_{st}(v)$ is the number of shortest paths going from
$s$ to $t$ passing through $v$ and $\sigma_{st}$ is the total
number of shortest paths going from $s$ to $t$. The end points of
each path is counted as part of the path\cite{Newman2001}. Newman
proposed a very fast algorithm taking only O($MN$) time to
calculate the betweenness of all vertices\cite{Newman2001}, where
$M$ and $N$ denote the number of edges and vertices, respectively.
The whole algorithm processes are as follows.

\begin{figure}
\scalebox{0.43}[0.43]{\includegraphics{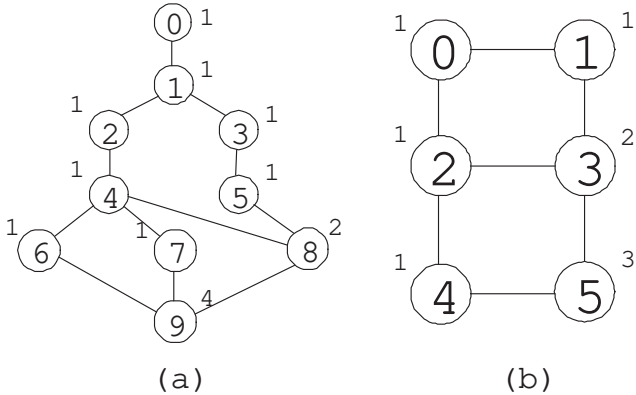}}
\caption{\label{fig:epsart} The two examples used to illuminate
the difference between Newman's and the corrected algorithms. (a)
The copy from Ref. [2], also bas been used as a sketch map for
Newman's algorithm. (b) The minimal network that can illuminate
the difference. The hollow circles represent the vertices and the
solid lines represent the edges. Each vertex is marked with a
natural number inside the corresponding circle, and the number
beside each vertex $v$ is $\sigma_{0v}$.}
\end{figure}

(1) Calculate the distance from a vertex $s$ to every other vertex
by using breadth-first search.

(2) A variable $b_v^s$, taking the initial value 1, is assigned to
each vertex $v$.

(3) Going through the vertices $v$ in order of their distance from
$s$, starting from the farthest, the value of $b_v^s$ is added to
corresponding variable on the predecessor vertex of $v$. If $v$
has more than one predecessor, then $b_v^s$ is divided {\bf
equally} between them.

(4) Go through all vertices in this fashion and records the value
$b_v^s$ for each $v$. Repeat the entire calculation for every
vertex $s$, the betweenness for each vertex $v$ is obtained as
\begin{equation}
B(v)=\sum_sb_v^s.
\end{equation}

\begin{table}
\caption{\label{tab:table1} Calculation results of figure 1(a)}
\begin{ruledtabular}
\begin{tabular}{@{}ccccccccccc@{}}
Vertices &
0 & 1 & 2 & 3 & 4 & 5 & 6 & 7 & 8 & 9 \\
Newman's & 9 & 34$\frac{5}{6}$ & 28$\frac{1}{6}$ & 22$\frac{1}{2}$ & 29$\frac{1}{3}$ & 21$\frac{2}{3}$ & 14$\frac{1}{4}$ & 14$\frac{1}{4}$ & 21$\frac{5}{6}$ & 24$\frac{1}{6}$ \\
Corrected & 9 & 34$\frac{1}{3}$ & 28$\frac{1}{3}$ & 21$\frac{2}{3}$ & 30 & 21 & 14$\frac{2}{3}$ & 14$\frac{2}{3}$ & 21$\frac{2}{3}$ & 24 \\
\end{tabular}
\end{ruledtabular}
\end{table}

\begin{table}
\caption{\label{tab:table2} Calculation results of figure 1(b)}
\begin{ruledtabular}
\begin{tabular}{@{}ccccccc@{}}
Vertices &
0 & 1 & 2 & 3 & 4 & 5 \\
Newman's & 6$\frac{3}{4}$ & 6$\frac{3}{4}$ & 11$\frac{1}{2}$ & 11$\frac{1}{2}$ & 6$\frac{3}{4}$ & 6$\frac{3}{4}$ \\
Corrected & 6$\frac{2}{3}$ & 6$\frac{2}{3}$ & 11$\frac{2}{3}$ & 11$\frac{2}{3}$ & 6$\frac{2}{3}$ & 6$\frac{2}{3}$ \\
\end{tabular}
\end{ruledtabular}
\end{table}

Since to a vertex $v$'s betweenness $B(v)$, the contributions of
its predecessors are not equal, it is not proper to divide $b_v^s$
equally between them. Clearly, if the vertex $v$ has $n$
predecessors labelled as $u_1,u_2,\cdots,u_n$ and $\sigma_{sv}$
different shortest paths to vertex $s$, then we have
\begin{equation}
\sigma_{sv}=\sum^n_{i=1}\sigma_{su_i}.
\end{equation}
The different shortest paths from $s$ to $v$ are divided into $n$
sets $G_1,G_2,\cdots,G_n$. The number of elements in $G_i$, that
is also the number of different shortest paths from $s$ to $u_i$ ,
gives expression to the contribution of the predecessor $u_i$ to
$v$'s betweenness. Therefore, the vertex $v$'s betweenness,
induced by the given source $s$, should be divided proportionally
to $\sigma_{su_i}$ rather than equally between its predecessors.
The corrected algorithm is as follows.

(1) Calculate the distance from a vertex $s$ to every other vertex
by using breadth-first search, taking time O($M$).

(2) Calculate the number of shortest paths from vertex $s$ to
every other vertex by using dynamic programming\cite{Bellman1962},
taking time O($M$) too. The processes are as follows. (2.1) Assign
$\sigma_{ss}=0$. (2.2) If all the vertices of distance $d(d\geq
0)$ is assigned (Note that the distance from $s$ to $s$ is zero),
then for each vertex $v$ whose distance is $d+1$, assign
$\sigma_{sv}=\sum_u\sigma_{su}$ where $u$ runs over all $v$'s
predecessors. (2.3) Repeat from step (2.1) until there are no
unassigned vertices left.

(3) A variable $\beta_v^s$, taking the initial value 1, is
assigned to each vertex $v$.

(4) Going through the vertices $v$ in order of their distance from
$s$, starting from the farthest, the value of $\beta_v^s$ is added
to corresponding variable on the predecessor vertex of $v$. If $v$
has more than one predecessor $u_1,u_2,\cdots,u_n$, $\beta_v^s$ is
multiplied by $\sigma_{su_i}/\sigma_{sv}$ and then added to
$\sigma_{su_i}$.

(5) Go through all vertices in this fashion and records the value
$\beta_v^s$ for each $v$. Repeat the entire calculation for every
vertex $s$, the betweenness for each vertex $v$ is obtained as
\begin{equation}
B(v)=\sum_s\beta_v^s.
\end{equation}

Clearly, the time complexity of the corrected algorithm is O($MN$)
too. Besides, one should pay attention to a more universal
algorithm proposed by Brandes\cite{Brandes2001}, which can be used
to calculate all kinds of centrality based on shortest-paths
counting for both unweighted and weighted networks.

These two algorithms, Newman's and the corrected one, will give
the same result if the network has a tree structure. However, when
the loops appear in the networks, the diversity between them can
be observed. Figure (1) exhibits two examples, the first one is
copied from the Ref. [2], and the second is the minimal network
that can illuminate the difference between Newman's and the
corrected algorithms. The comparisons between these two algorithms
are shown in table (1) and (2). The two algorithms produce
different results even for networks of very few vertices.

\begin{figure}
\scalebox{1}[1]{\includegraphics{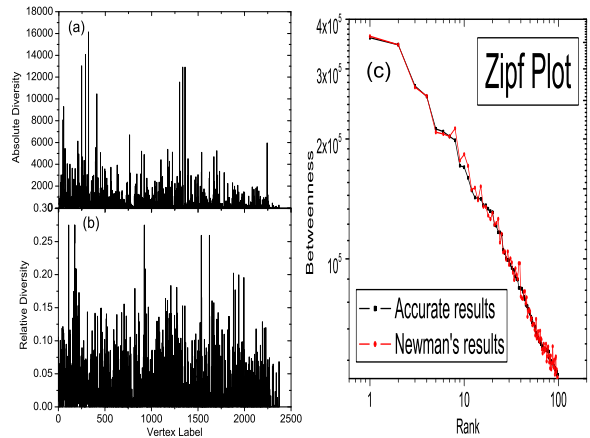}}
\caption{\label{fig:epsart} The comparisons between Newman's and
the corrected algorithms on the protein interaction network of
Yeast. (a) and (b) shown the absolute diversity and relative
diversity between Newman's and the accurate results, respectively.
(c) is the Zipf plot of the 100 vertices with highest
betweenness.}
\end{figure}

In addition, we compare with the performances of these two
algorithms on the protein interaction network of
Yeast\cite{Jeong2001}. This network has 2617 vertices, but only
its maximal component containing 2375 vertices is taken into
account. Figure 2(a) and 2(b) report the absolute diversity and
relative diversity between Newman's and the accurate (obtained
from the corrected algorithm) results, respectively. The departure
is distinct and can not be neglected. Fortunately, the statistical
features may be similar. Although the details of the Zipf
plot\cite{Zipf1949} of the top-100 vertices are not the same, both
the two curves obey power-law form with almost the same exponent.
We also have checked that the scaling law\cite{Goh2001,Goh2002} of
betweenness distribution in Barab\'{a}si-Albert
networks\cite{Barabasi1999} is kept, while the power-law exponents
are slightly changed.

The measure of betweenness is now widely used to detect
communities/modules structures\cite{Girvan2002,Newman2004} and to
analysis dynamics upon networks. Since the statistical characters
of betweenness distributions obtained by Newman's and the
corrected algorithm are almost the same, some researchers may have
found the difference between these two algorithm but have not paid
attention to it. However, many previous works have demonstrated
that a few nodes' betweennesses rather than the overall
betweenness distribution, may sometimes, determine the key
features of dynamic behaviors on networks. Examples are numerous:
these include the network
traffics\cite{Guimera2002,Zhao2005,Yan2005}, the
synchronization\cite{Nishikawa2003,Hong2004,ZhaoM2005}, the
cascading dynamics\cite{Motter2004}, and so on. In figure 2(b),
one can find that for many nodes the relative diversities
betweenness those two algorithms exceed $10\%$, and even nearly
$30\%$ for a few nodes. Therefore, the difference can not be
neglected especially in analyzing the networks dynamics.

Although Newman's algorithm does not agree with the definition of
betweenness\cite{Newman2001}, it may be more practical especially
for the large-scale communication systems wherein the routers do
not know how many shortest paths there are to the destination.
Even if they can save the information of all the successors'
weights, to implement the biased choices may bring additional
costs in economy and technique. Hence just to choose with equal
probability at each branch point may be more natural, which is in
accordance with Newman's algorithm.


\begin{thebibliography} {Freeman1977}

\bibitem{Anthonisse1971} J. M. Anthonisse, Technical Report BN 9/71, Stichting Mathematisch Centrum, Amsterdam.
\bibitem{Freeman1977} L. C. Freeman, Sociometry {\bf 40}, 35 (1977).
\bibitem{Newman2001} M. E. J. Newman, Phys. Rev. E {\bf 64}, 016132 (2001).
\bibitem{Bellman1962} R. E. Bellman, and S. E. Dreyfus, {\it Applied Dynamic Programming} (Princeton University Press, New Jersy, 1962).
\bibitem{Brandes2001} U. Brandes, Journal of Mathematical Sociology {\bf 25}, 163(2001).
\bibitem{Jeong2001} H. Jeong, S. Mason, A. -L. Barab\'{a}si, and Z. N. Oltvai, Nature {\bf 411}, 41 (2001).
\bibitem{Zipf1949} G. K. Zipf, {\it Human Behavior and the Principal of Least Effort} (Addison-Wesley, Cambridge, MA, 1949).
\bibitem{Goh2001} K. -I. Goh, B. Kahng, and D. Kim, Phys. Rev. Lett. {\bf 87}, 278701 (2001).
\bibitem{Goh2002} K. -I. Goh, E. Oh, H. Jeong, B. Kahng, and D. Kim, Proc. Natl. Acad. Sci. USA {\bf 99}, 12583 (2002).
\bibitem{Barabasi1999} A. -L. Barab\'{a}si, and R. Albert, Science {\bf 286}, 509 (1999).
\bibitem{Girvan2002} M. Girvan, and M. E. J. Newman, Proc. Natl. Acad. Sci. U.S.A. {\bf 99}, 7821 (2002).
\bibitem{Newman2004} M. E. J. Newman, and M. Girvan, Phys. Rev. E {\bf 69}, 026113(2004).
\bibitem{Guimera2002} R. Guimer\'a, A. D\'iaz-Guilera, F. Vega-Redondo, A. Cabrales, and A. Arenas, Phys. Rev. Lett. \textbf{89}, 248701 (2002).
\bibitem{Zhao2005} L. Zhao, Y.-C. Lai, K. Park, and N. Ye, Phys. Rev. E \textbf{71}, 026125 (2005).
\bibitem{Yan2005} G. Yan, T. Zhou, B. Hu, Z. -Q. Fu, and B. -H. Wang, arXiv: cond-mat/0505366.
\bibitem{Nishikawa2003} T. Nishikawa, A. E. Motter, Y. -C. Lai, and F. C. Hoppensteadt, Phys. Rev. Lett. {\bf 91} 014101 (2003).
\bibitem{Hong2004} H. Hong, B. J. Kim, M. Y. Choi, and H. Park, Phys. Rev. E {\bf 69}, 067105(2004).
\bibitem{ZhaoM2005} M. Zhao, T. Zhou, B. -H. Wang, and W. -X. Wang, Phys. Rev. E {\bf 72}, 057102(2005).
\bibitem{Motter2004} A. E. Motter, Phys. Rev. Lett. {\bf 93}, 098701 (2004).

\end{thebibliography}
\end{document}